\documentclass[reprint,amsmath,amssymb,superscriptaddress]{revtex4-1}

\usepackage{graphicx}
\usepackage{dcolumn}
\usepackage{bm}
\usepackage{color}
\usepackage{ulem}
\begin{document}

%\preprint{}

\title{Interface-driven giant tunnel magnetoresistance in (111)-oriented junctions}


\author{Keisuke Masuda}
%\email{MASUDA.Keisuke@nims.go.jp}
\affiliation{Research Center for Magnetic and Spintronic Materials, National Institute for Materials Science (NIMS), Tsukuba 305-0047, Japan}
\author{Hiroyoshi Itoh}
%\email{hitoh@kansai-u.ac.jp}
\affiliation{Department of Pure and Applied Physics, Kansai University, Suita 564-8680, Japan}
\affiliation{Center for Spintronics Research Network, Osaka University, Toyonaka 560-8531, Japan}
\author{Yoshio Miura}
%\email{MIURA.Yoshio@nims.go.jp}
\affiliation{Research Center for Magnetic and Spintronic Materials, National Institute for Materials Science (NIMS), Tsukuba 305-0047, Japan}
\affiliation{Center for Spintronics Research Network, Osaka University, Toyonaka 560-8531, Japan}
\affiliation{Center for Materials Research by Information Integration, National Institute for Materials Science (NIMS), Tsukuba 305-0047, Japan}
%\thanks{}
%\affiliation{}


\date{\today}

\begin{abstract}
We theoretically study the tunnel magnetoresistance (TMR) effect in (111)-oriented junctions Co/MgO/Co(111) and Ni/MgO/Ni(111). The Co-based junction is shown to have a TMR ratio over 2000$\%$, which is one order higher than that of the Ni-based one. The high TMR ratio is attributed to the interfacial resonance effect: The interfacial $d$--$p$ antibonding states are formed close to the Fermi level in the majority-spin channel and these states in both interfaces resonate with each other. This differs essentially from the conventional coherent tunneling mechanism of high TMR ratios in Fe(Co)/MgO/Fe(Co)(001).
\end{abstract}

%\keywords{}

\maketitle

\section{\label{introduction} introduction}
\begin{figure}
\includegraphics[width=8.7cm]{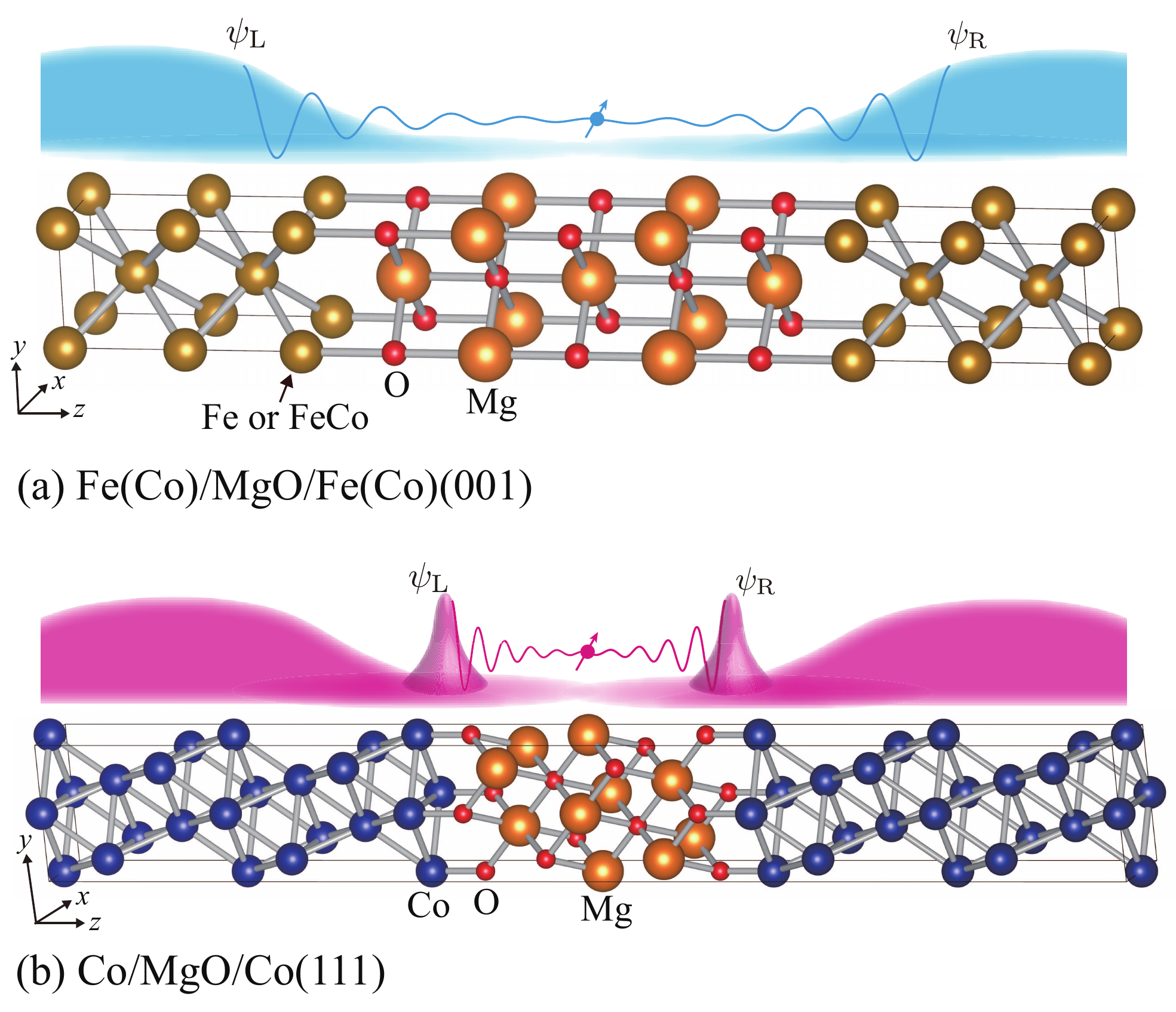}
\caption{\label{fig1} Supercells of (a) Fe(Co)/MgO/Fe(Co)(001) and (b) Co/MgO/Co(111). Schematics for the TMR effects in these MTJs are also shown.}
\end{figure}

Since the observation of the giant tunnel magnetoresistance (TMR) effect in Fe(Co)/MgO/Fe(Co)(001) magnetic tunnel junctions (MTJs) \cite{2004Yuasa-NatMat,2004Parkin-NatMat}, the TMR effect has long been explained by the coherent tunneling mechanism \cite{2001Butler-PRB,2001Mathon-PRB}: Bulk wave functions of the ferromagnetic electrode are selectively filtered by the MgO barrier and only the $\Delta_1$ wave function with half metallicity at the Fermi level passes through the barrier, leading to the high TMR ratio [Fig. \ref{fig1}(a)]. However, there is a significant discrepancy in the TMR ratio between the theory and experiments; the highest TMR ratio observed so far is around 500\% at low temperature \cite{2006Yuasa-APL}, but is about half of the theoretically predicted value over 1000\% \cite{2001Butler-PRB,2001Mathon-PRB}. A possible key to understand this gap is interfacial effects. Several studies \cite{2011Miura-PRB,2009Sakuma-JAP} have indicated the importance of interfacial states for explaining temperature dependencies of TMR ratios \cite{2008Tsunegi-APL}. In the (001)-oriented MTJs, the interfacial states are formed in the minority-spin state \cite{2001Butler-PRB,2008Kim-JAP} and tend to decrease the TMR ratio \cite{1975Julliere-PL}.

These motivate us to speculate that interfacial states provide significant contribution to TMR effects in real experiments and decrease the TMR ratios in Fe(Co)/MgO/Fe(Co)(001) MTJs. In contrast, we can utilize such interfacial states for enhancing the TMR effect significantly; this study proposes a quite high TMR ratio driven by the interfacial resonance effect in an unconventional (111)-oriented MTJ.

In the present work, we theoretically examine the TMR effect in two basic (111)-oriented MTJs, Co/MgO/Co(111) [Fig. \ref{fig1}(b)] and Ni/MgO/Ni(111), where (111) directions of the MgO barrier and ferromagnetic electrodes (fcc Co or fcc Ni) are parallel to the stacking direction of the MTJ. It is natural to consider such (111)-oriented MTJs for fcc ferromagnetic electrodes because the close-packed (111) plane has the lowest surface energy in the fcc lattice \cite{1988Ting-SurfSci}. However, TMR effects in these MTJs have not been understood well, since most previous studies have focused on (001)-oriented MTJs with bcc ferromagnetic electrodes. We calculate conductances and TMR ratios of the (111)-oriented MTJs by means of the first-principles approach combining the density-functional theory (DFT) and the Landauer formula. It is shown that the obtained TMR ratio of the Co-based MTJ is quite high ($\sim 2100 \%$) while that of the Ni-based one is relatively low ($\sim 250 \%$). From the in-plane wave-vector dependencies of the conductances, we find that the TMR effect in the (111)-oriented MTJs cannot be understood from the bulk band structures of the barrier and electrodes, which is essentially different from the case of the (001)-oriented MTJs. Detailed analyses of the local electronic structure and the transmittance clarify that the resonance of the interfacial $d$--$p$ antibonding states in the majority-spin channel is the origin of the high TMR ratio in the Co-based (111)-oriented MTJ.

\section{\label{method} calculation method}
We prepared supercells of Co/MgO/Co(111) and Ni/MgO/Ni(111) as shown in Fig. \ref{fig1}(b). The $a$- and $b$-axis lengths were fixed to $a=a_{\rm fcc}/\sqrt{2}$ and $b=\sqrt{3}\,a_{\rm fcc}/\sqrt{2}$, where we used $a_{\rm fcc}=3.52\,{\rm \AA}$ for both the supercells. In the Co-based (Ni-based) supercell, we chose the Co-O (Ni-O) interface, which is energetically favored compared to the Co-Mg (Ni-Mg) one. The atomic positions along the $c$ direction in the supercells were relaxed using the first-principles DFT calculation implemented in the Vienna {\it ab initio} simulation program ({\scriptsize VASP}) \cite{1996Kresse-PRB}. Here, we adopted the spin-polarized generalized gradient approximation (GGA) \cite{1996Perdew-PRL} for the exchange-correlation energy and used the projector augmented wave pseudopotential \cite{1994Bloechl-PRB,1999Kresse-PRB} to treat the effect of core electrons properly. A cutoff energy of 500\,eV was employed and the Brillouin-zone integration was performed with the $23\times13\times1$ Monkhorst-Pack ${\bf k}$-point grid. The convergence criteria for energy and force were set to $10^{-5}\,{\rm eV}$ and $10^{-3}\,{\rm eV/\AA}$, respectively. More details of our structure optimization are given in our previous work \cite{2017Masuda-PRB_bias}.

Using the optimized supercell, we constructed quantum open systems by attaching the left and right semi-infinite electrodes of Co (Ni) to the Co-based (Ni-based) supercell. In each quantum open system, we calculated conductances for both parallel and antiparallel configurations of magnetization in the electrodes using the {\scriptsize PWCOND} code \cite{2004Smogunov-PRB} in the {\scriptsize QUANTUM ESPRESSO} package \cite{Baroni}. First, we obtained the self-consistent potential of the quantum open system, where the GGA and the ultrasoft pseudopotentials were used for the DFT calculation. The cutoff energies for the wave functions and the charge density were fixed to 45 and 450 Ry, respectively. The 23$\times$13$\times$1 ${\bf k}$ points were used for the Brillouin-zone integration and the convergence criterion was set to $10^{-6}\,{\rm Ry}$. Since our systems are repeated periodically in the $xy$-plane, the scattering states can be classified by an in-plane wave vector ${\bf k}_\parallel =(k_x,k_y)$. For each ${\bf k}_\parallel$ and spin index, we solved the scattering equation derived under the condition that the wave function and its derivative of the supercell are connected to those of the electrodes \cite{2004Smogunov-PRB,1999Choi-PRB}. These calculations and the Landauer formula give the wave-vector-resolved conductances $G_{{\rm P},\uparrow}({\bf k}_\parallel)$, $G_{{\rm P},\downarrow}({\bf k}_\parallel)$, $G_{{\rm AP},\uparrow}({\bf k}_\parallel)$, and $G_{{\rm AP},\downarrow}({\bf k}_\parallel)$, which are the majority- and minority-spin conductances in the parallel and antiparallel configurations of magnetizations, respectively. The averaged conductances are obtained as, e.g., $G_{{\rm P},\uparrow}=\sum_{{\bf k}_\parallel}G_{{\rm P},\uparrow}({\bf k}_\parallel)/N$, where $N$ is the sampling number of ${\bf k}_\parallel$ points. In the present study, $N$ was set to 2500 ensuring good convergence for the conductances. Using the averaged conductances, we calculated ${\rm TMR\,\, ratio}\,(\%)=100\times(G_{\rm P}-G_{\rm AP})/G_{\rm AP}$, where $G_{\rm P(AP)}=G_{{\rm P(AP)},\uparrow}+G_{{\rm P(AP)},\downarrow}$.

\section{\label{resultsdiscussion} results and discussion}
\begin{table}[t]
\caption{\label{tab1}
Calculated conductances and TMR ratios. The units are in $e^2\!/h$ and $\%$, respectively.
}
\begin{ruledtabular}
\begin{tabular}{ccc}
\textrm{}&
\textrm{Co/MgO/Co(111)}&
\textrm{Ni/MgO/Ni(111)}\\
\colrule
$G_{{\rm P},\uparrow}$ & 2.48$\times$10$^{-3}$ & 5.71$\times$10$^{-5}$\\
$G_{{\rm P},\downarrow}$ & 1.42$\times$10$^{-3}$ & 1.23$\times$10$^{-3}$\\
$G_{{\rm AP},\uparrow}$ & 8.73$\times$10$^{-5}$ & 1.84$\times$10$^{-4}$\\
$G_{{\rm AP},\downarrow}$ & 8.75$\times$10$^{-5}$ & 1.84$\times$10$^{-4}$\\
$G_{\rm P}$ & 3.90$\times$10$^{-3}$ & 1.29$\times$10$^{-3}$\\
$G_{\rm AP}$ & 1.75$\times$10$^{-4}$ & 3.68$\times$10$^{-4}$\\
TMR ratio & 2130 & 250\\
\end{tabular}
\end{ruledtabular}
\end{table}
Table \ref{tab1} shows conductances and TMR ratios obtained in our calculations. In the Co-based MTJ, the parallel conductance $G_{\rm P}$ is much larger than the antiparallel one $G_{\rm AP}$, leading to a quite high TMR ratio of more than 2000$\%$. On the other hand, the Ni-based MTJ has a smaller TMR ratio of 250$\%$ since the difference between $G_{\rm P}$ and $G_{\rm AP}$ is smaller compared to the Co-based case. One may think that the mechanism of such TMR effects is similar to that in the (001)-oriented MTJs. Previous theoretical studies \cite{2001Butler-PRB,2001Mathon-PRB} have shown that the TMR effect in Fe/MgO/Fe(001) is dominated by the bulk band structures of Fe and MgO along the $\Delta$ line corresponding to the (001) direction; in their results, $G_{{\rm P},\uparrow}({\bf k}_\parallel)$ mainly contributing to the high TMR ratio has a sharp peak at ${\bf k}_\parallel=(0,0)=\Gamma$. If the similar mechanism holds for the (111)-oriented MTJs, the TMR effect should be explained by the band structures along the $\Lambda$ line corresponding to the (111) direction. In this case, $G_{{\rm P},\uparrow}({\bf k}_\parallel)$ or $G_{{\rm P},\downarrow}({\bf k}_\parallel)$ or both of them should have a large value at ${\bf k}_\parallel=\Gamma$, since the $\Lambda$ line is equivalent to the $k_z$ line at ${\bf k}_\parallel=\Gamma$ in the present supercells [Fig. \ref{fig1}(b)].

\begin{figure}
\includegraphics[width=8.5cm]{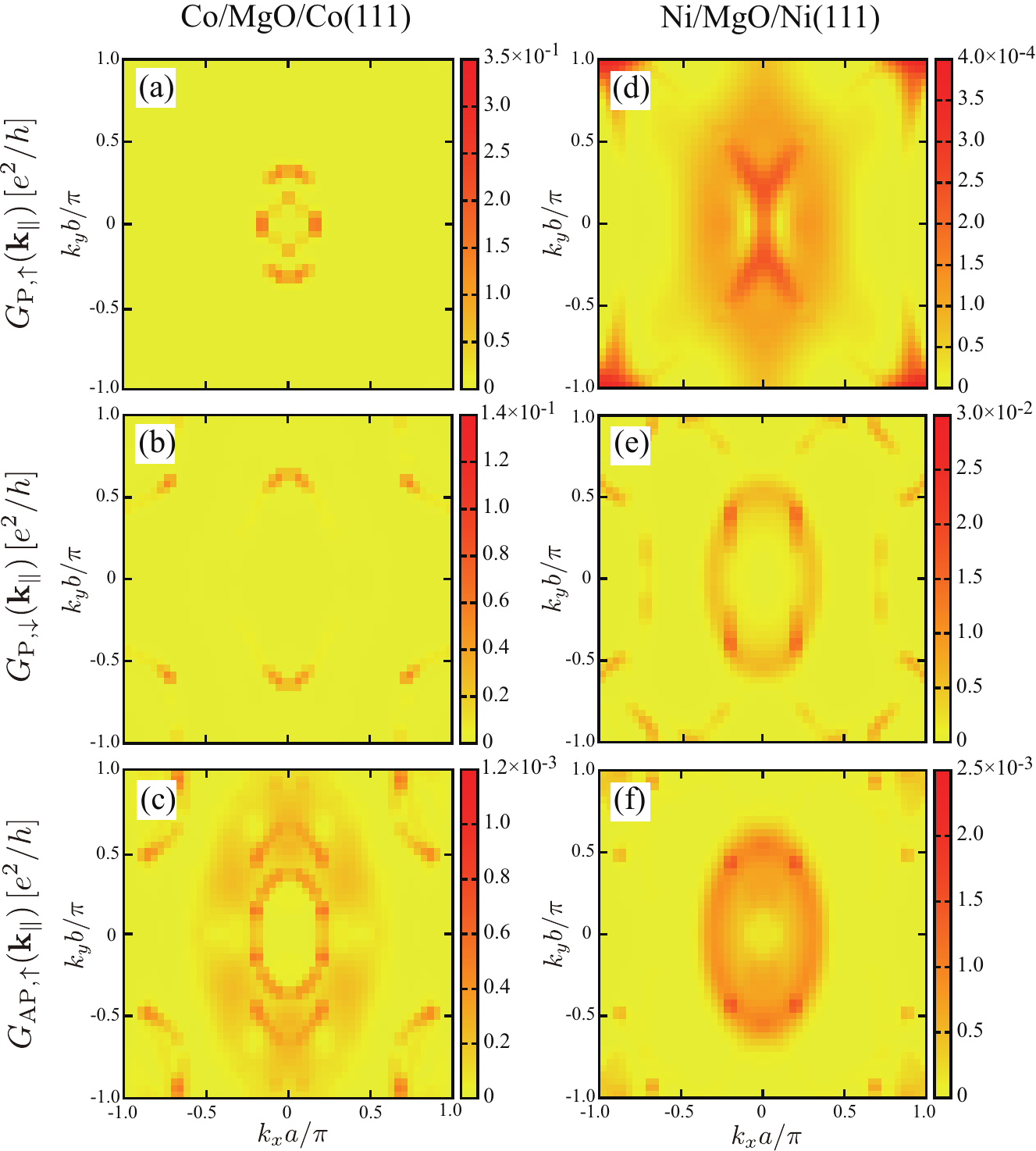}
\caption{\label{kpara-dep} The ${\bf k}_\parallel$ dependencies of conductances in Co/MgO/Co(111) [(a)-(c)] and Ni/MgO/Ni(111) [(d)-(f)]. (a),(d) Majority-spin conductances $G_{{\rm P},\uparrow}({\bf k}_\parallel)$ and (b),(e) minority-spin conductances $G_{{\rm P},\downarrow}({\bf k}_\parallel)$ in the parallel magnetization configurations. (c),(f) Majority-spin conductances $G_{{\rm AP},\uparrow}({\bf k}_\parallel)$ in the antiparallel magnetization configurations.}
\end{figure}
\begin{figure}
\includegraphics[width=8.8cm]{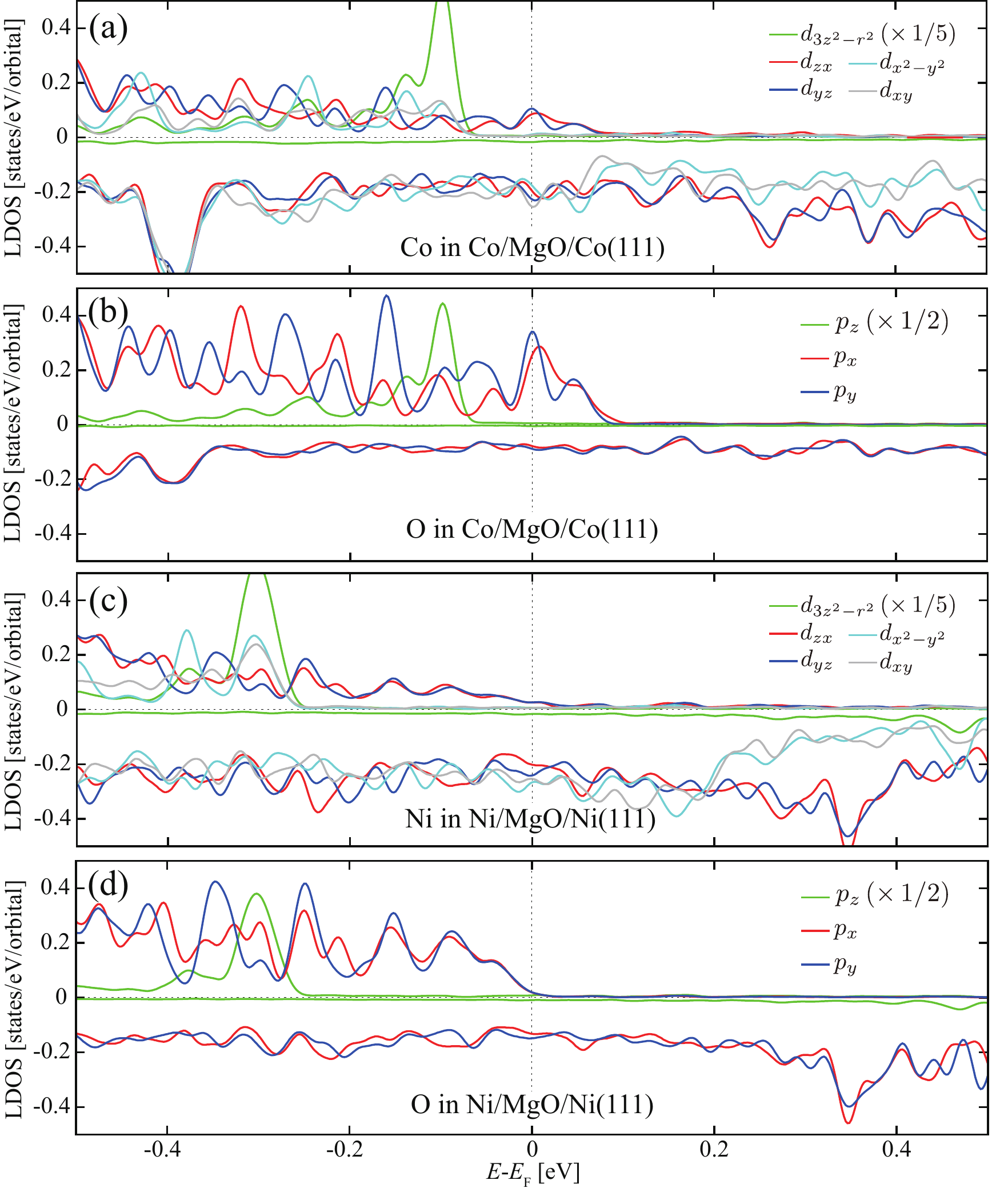}
\caption{\label{interfacial-LDOS} (a),(b) Projected LDOSs at interfacial Co and O atoms in Co/MgO/Co(111). (c),(d) The same as (a) and (b) but at interfacial Ni and O atoms in Ni/MgO/Ni(111). In each panel, positive and negative values indicate the majority- and minority-spin projected LDOSs.}
\end{figure}
To see whether or not this scenario is valid, we analyzed the in-plane wave vector ${\bf k}_{\parallel}=(k_x,k_y)$ dependencies of the conductances as shown in Fig. \ref{kpara-dep}. Figure \ref{kpara-dep}(a) shows the majority-spin conductance $G_{{\rm P},\uparrow}({\bf k}_{\parallel})$ in the Co-based MTJ playing the key role for the obtained high TMR ratio. We see that the conductance at ${\bf k}_{\parallel}=\Gamma$ is much smaller than that around ${\bf k}_{\parallel}=\Gamma$, which is clearly different from the features expected from the above-mentioned mechanism. Figures \ref{kpara-dep}(b) and \ref{kpara-dep}(c) show the minority-spin and antiparallel conductances [$G_{{\rm P},\downarrow}({\bf k}_{\parallel})$ and $G_{{\rm AP},\uparrow}({\bf k}_{\parallel})$] in the Co-based MTJ \cite{remark_AP-conductance}, which also do not have a significant value at ${\bf k}_{\parallel}=\Gamma$ and their values spread more widely in the ${\bf k}_{\parallel}$ Brillouin zone than $G_{{\rm P},\uparrow}({\bf k}_{\parallel})$. In Figs. \ref{kpara-dep}(d)--\ref{kpara-dep}(f), we show the ${\bf k}_{\parallel}$ dependencies of the conductances in the Ni-based MTJ. Although detailed features are different from those in the Co-based MTJ, the conductances also do not have a significant value at ${\bf k}_{\parallel}=\Gamma$. All these results indicate that the TMR effect in the (111)-oriented MTJs cannot be explained by the bulk band structures of the barrier and electrodes along the $\Lambda$ line, in sharp contrast to the previous results on the (001)-oriented MTJs \cite{2001Butler-PRB,2001Mathon-PRB}. We also analyzed bulk band structures of fcc Co, Ni, and MgO along the $\Lambda$ line, which revealed that, although the complex band of MgO(111) consists of the $\Lambda_1$ state, both Co and Ni do not have a half metallicity in the $\Lambda_1$ state. This also suggests the inapplicability of the explanation based on bulk band structures.

To obtain further insight into the role of the MgO barrier, we additionally analyzed the TMR effect in Co/vacuum/Co(111), where the distance between the left and right interfaces and all the positions of Co atoms were set to be the same as those of Co/MgO/Co(111). We obtained a TMR ratio of 320\%, which is one order lower than that of Co/MgO/Co(111). This clearly indicates the necessity of MgO to achieve a high TMR ratio. From the comparison of the ${\bf k}_{\parallel}$-resolved conductances between the vacuum and MgO cases, we found that MgO particularly damps the wave functions that are distributed away from ${\bf k}_{\parallel}=\Gamma$.

\begin{figure}
\includegraphics[width=8.8cm]{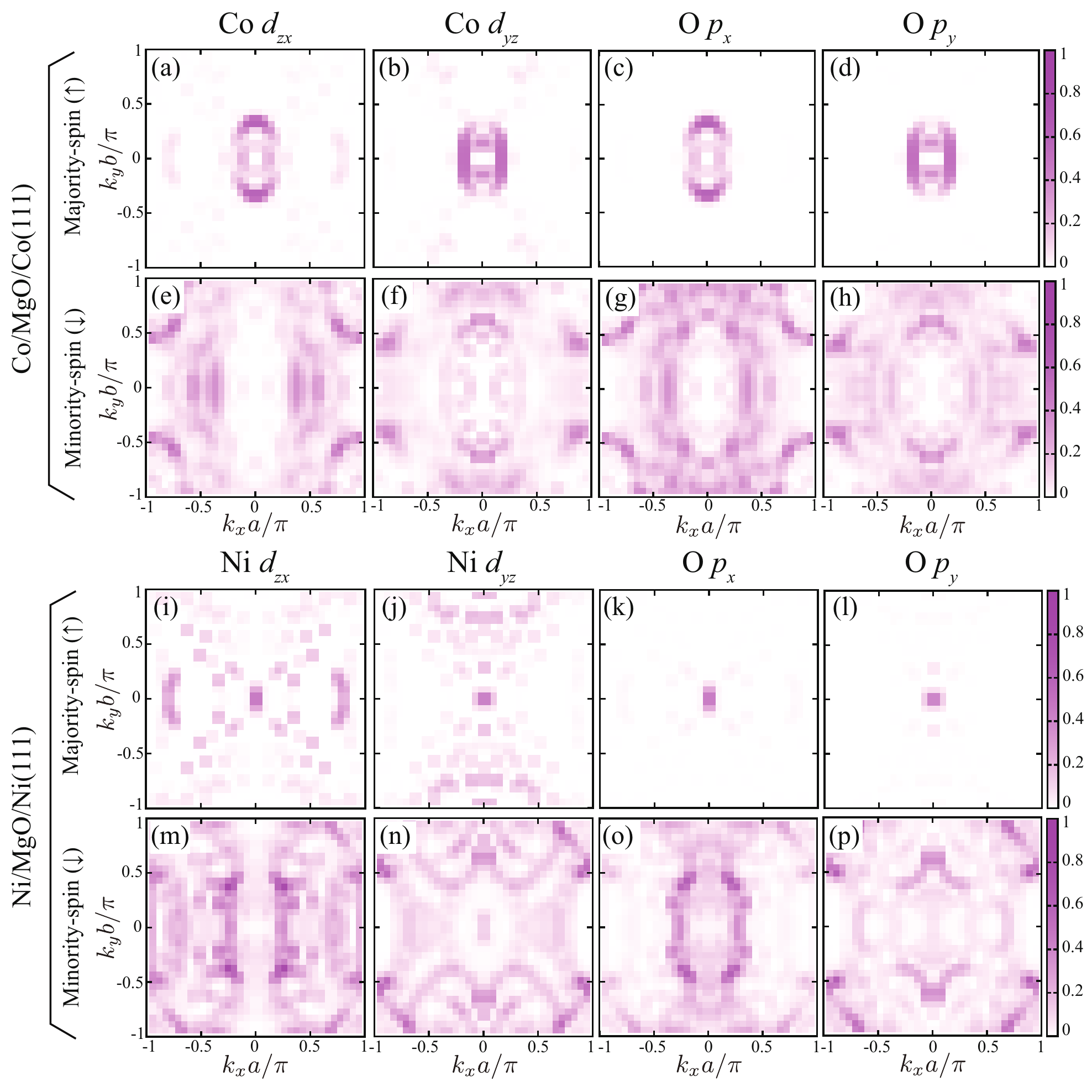}
\caption{\label{kresolved-LDOS} The ${\bf k}_{\parallel}$-resolved LDOSs at $E=E_{\rm F}$ of interfacial atoms in (a)-(h) Co/MgO/Co(111) and (i)-(p) Ni/MgO/Ni(111). (a)-(d) Contributions from Co $d_{zx}$, Co $d_{yz}$, O $p_{x}$, and O $p_{y}$ orbitals in the majority-spin state, respectively. (e)-(h) The same as (a)-(d) but in the minority-spin state. (i)-(l) Contributions from Ni $d_{zx}$, Ni $d_{yz}$, O $p_{x}$, and O $p_{y}$ orbitals in the majority-spin state, respectively. (m)-(p) The same as (i)-(l) but in the minority-spin state. In each panel, ${\bf k}_{\parallel}$-resolved LDOSs are normalized by its maximum value.}
\end{figure}
Interfacial effects are the key to understand the present TMR effect. As shown in Figs. \ref{kpara-dep}(a)--\ref{kpara-dep}(f), the conductances have significant values in a set of ${\bf k}_{\parallel}$ points surrounding ${\bf k}_{\parallel}=\Gamma$. This reminds us of the existence of the interfacial resonance effect. Examples showing this effect are Fe/MgO/Fe(001) \cite{2001Butler-PRB,2006Waldron-PRL,2007Rungger-JMMM,2009Rungger-PRB}, Co/MgO/Co(001) \cite{2004Zhang-PRB}, and FeCo/MgO/FeCo(001) \cite{2004Zhang-PRB}, which have interfacial states close to the Fermi level in the minority-spin channel. Such states in the left and right interfaces resonate with each other and provide non-negligible values of $G_{{\rm P},\downarrow}({\bf k}_{\parallel})$ in the ${\bf k}_{\parallel}$ points surrounding $\Gamma$. Furthermore, additional transport between the interfacial minority-spin states and the bulk majority-spin $\Delta_1$ state enhance $G_{{\rm AP},\uparrow}({\bf k}_{\parallel})$ and $G_{{\rm AP},\downarrow}({\bf k}_{\parallel})$. Thus, the interfacial states itself tend to decrease the TMR ratio in the (001)-oriented MTJ. However, since $G_{{\rm P},\uparrow}({\bf k}_{\parallel})$ has a quite large value at ${\bf k}_{\parallel}=\Gamma$ based on the coherent tunneling mechanism \cite{2001Butler-PRB,2001Mathon-PRB}, the interfacial effect provides a small contribution to the TMR ratio. Note that these are the previous theoretical results; the interfacial effect is sensitive to the interfacial conditions in real experiments \cite{2006Waldron-PRL}.

To confirm the existence of interfacial states in the present MTJs, we show in Figs. \ref{interfacial-LDOS}(a) and \ref{interfacial-LDOS}(b) the projected LDOSs of interfacial Co and O atoms in the Co/MgO/Co(111) MTJ. In Fig. \ref{interfacial-LDOS}(b), we see that the interfacial O atoms have peaks of the majority-spin LDOSs in the $p_x$ and $p_y$ orbitals close to the Fermi level ($E_{\rm F}$), which originates from the antibonding between Co $d_{zx}$ ($d_{yz}$) and O $p_x$ ($p_y$) states at the interface. As mentioned below, such interfacial antibonding states mainly contribute to the high TMR ratio through the interfacial resonance effect. It should be emphasized that these majority-spin interfacial states enhance $G_{{\rm P},\uparrow}({\bf k}_{\parallel})$ but hardly enhance $G_{{\rm AP},\uparrow}({\bf k}_{\parallel})$ and $G_{{\rm AP},\downarrow}({\bf k}_{\parallel})$, which is because the bulk minority-spin state is not significant in the (111)-oriented MTJ. This is clearly different from the (001)-oriented case and the reason why we obtained the large TMR ratio in this MTJ. In the case of the Ni/MgO/Ni(111) MTJ [Figs. \ref{interfacial-LDOS}(c) and \ref{interfacial-LDOS}(d)], on the other hand, the interfacial antibonding state between Ni $d_{zx}$ ($d_{yz}$) and O $p_x$ ($p_y$) states is formed at a lower energy compared to the Co-based case, which is owing to the difference in the valence electron number between Co and Ni. Thus, the interfacial O atoms have small majority-spin LDOSs in the $p_x$ and $p_y$ orbitals at $E=E_{\rm F}$.

Conclusive information on the mechanism for the present TMR effect is given by the ${\bf k}_{\parallel}$-resolved LDOSs of interfacial atoms in Co/MgO/Co(111) [Figs. \ref{kresolved-LDOS}(a)--\ref{kresolved-LDOS}(h)] and Ni/MgO/Ni(111) [Figs. \ref{kresolved-LDOS}(i)--\ref{kresolved-LDOS}(p)]. Here, we only showed the LDOSs at $E=E_{\rm F}$ in the Co (Ni) $d_{zx}$, Co (Ni) $d_{yz}$, O $p_x$, and O $p_y$ orbitals providing essential contributions to the conductances. It is seen that the LDOS in the Co (Ni) $d_{zx}$ orbital has an almost the same ${\bf k}_{\parallel}$ dependence as that in the O $p_x$ orbital in each spin state [e.g., Figs. \ref{kresolved-LDOS}(a) and \ref{kresolved-LDOS}(c)], since these orbitals make an antibonding state around $E=E_{\rm F}$ as mentioned above. The same relation holds between the LDOSs of Co (Ni) $d_{yz}$ and O $p_y$ orbitals [e.g., Figs. \ref{kresolved-LDOS}(b) and \ref{kresolved-LDOS}(d)]. Of particular importance is that the ${\bf k}_{\parallel}$ dependencies of the conductances in Fig. \ref{kpara-dep} can be understood by those of the LDOSs in Fig. \ref{kresolved-LDOS}. In fact, $G_{{\rm P},\uparrow}({\bf k}_{\parallel})$ in the Co-based MTJ [Fig. \ref{kpara-dep}(a)] can be almost reproduced by mixing the ${\bf k}_{\parallel}$ dependencies of majority-spin LDOSs in the O $p_x$ (Co $d_{zx}$) and O $p_y$ (Co $d_{yz}$) states [Figs. \ref{kresolved-LDOS}(a)--\ref{kresolved-LDOS}(d)]. The minority-spin conductance $G_{{\rm P},\downarrow}({\bf k}_{\parallel})$ [Fig. \ref{kpara-dep}(b)] also reflects minority-spin LDOSs in the O $p_x$ (Co $d_{zx}$) and O $p_y$ (Co $d_{yz}$) states [Figs. \ref{kresolved-LDOS}(e)--\ref{kresolved-LDOS}(h)]. In a similar way, we can explain the ${\bf k}_{\parallel}$ dependencies of the conductances [Figs. \ref{kpara-dep}(d)-\ref{kpara-dep}(f)] by those of the LDOSs [Figs. \ref{kresolved-LDOS}(i)--\ref{kresolved-LDOS}(p)] in the case of Ni/MgO/Ni(111); e.g., $G_{{\rm P},\downarrow}({\bf k}_{\parallel})$ [Fig. \ref{kpara-dep}(e)] has a similar ${\bf k}_{\parallel}$ dependence to the minority-spin LDOSs in the Ni $d_{zx}$ and O $p_x$ states [Figs. \ref{kresolved-LDOS}(m) and \ref{kresolved-LDOS}(o)]. All these results clearly suggest that the interfacial antibonding states at $E \approx E_{\rm F}$ provide the TMR effect in the (111)-oriented MTJs through the interfacial resonance effect. 

\begin{figure}
\includegraphics[width=8.5cm]{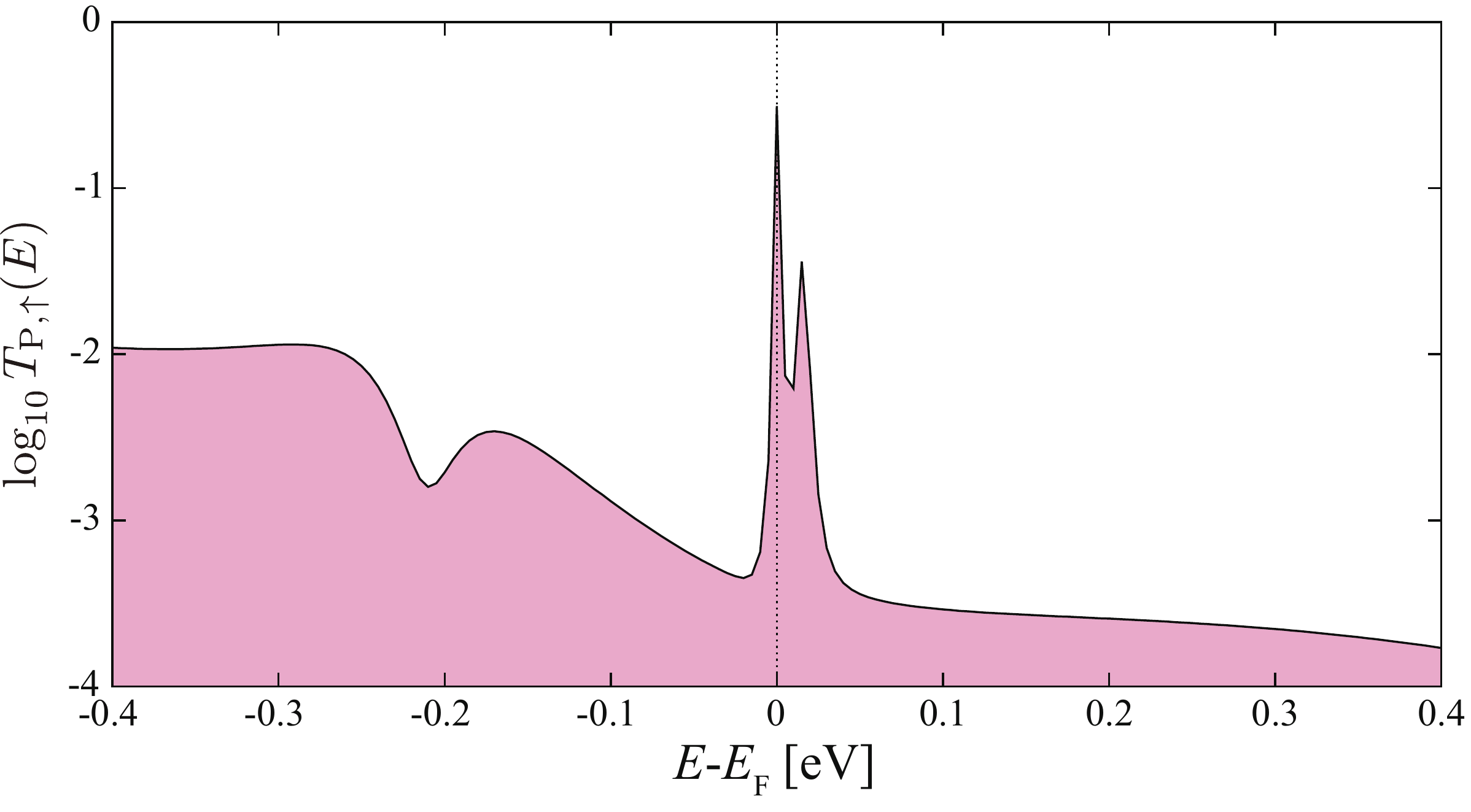}
\caption{\label{edep} The energy dependence of the majority-spin transmittance at $(k_x a/\pi,k_y b/\pi)=(0.04,0.32)$ in the Co/MgO/Co(111) MTJ with the parallel configuration of magnetizations.}
\end{figure}
Further evidence for the interfacial resonance effect is obtained from the energy dependence of the transmittance. Figure \ref{edep} shows the majority-spin transmittance $T_{{\rm P},\uparrow}(E)$ of the Co/MgO/Co(111) MTJ at $(k_x a/\pi,k_y b/\pi)=(0.04,0.32)$, where the conductance $G_{{\rm P},\uparrow}({\bf k}_{\parallel})$ has the maximum value [Fig. \ref{kpara-dep}(a)]. Note that the transmittance is converted to the conductance simply by multiplying $e^2/h$ [$G_{{\rm P},\uparrow}({\bf k}_{\parallel})=(e^2/h) \times T_{{\rm P},\uparrow}({\bf k}_{\parallel})$] and we showed in Fig. \ref{kpara-dep} the conductances at $E=E_{\rm F}$. In Fig. \ref{edep}, we can see sharp peaks close to $E_{\rm F}$, which are similar to those previously obtained in the minority-spin transmittance in the Fe/MgO/Fe(001) MTJ exhibiting the interfacial resonance effect \cite{2006Waldron-PRL,2007Rungger-JMMM,2009Rungger-PRB}. Since the resonance effect usually occurs in a narrow energy range, such a characteristic energy dependence of the transmittance is concluded to originate from the interfacial resonance effect. On the other hand, when a finite bias voltage ($\sim 0.05\,{\rm V}$) is applied to the MTJ, the interfacial resonance could be removed owing to its small energy width \cite{2006Waldron-PRL,2002Wunnicke-PRB}, leading to the reduction of the TMR ratio. However, we have many other interfacial states around $E_{\rm F}$ as shown in Figs. \ref{interfacial-LDOS}(a) and \ref{interfacial-LDOS}(b). Thus, if we increase the bias voltage further, these interfacial states may form a new resonance and enhance the TMR ratio \cite{2006Waldron-PRL}.

Some previous studies have provided partial information on (111)-oriented MTJs. Hauch {\it et al.} \cite{2008Hauch-APL} fabricated an Fe(110)/MgO(111)/Fe(110) MTJ with the (111)-oriented MgO barrier and observed a low TMR ratio of 54$\%$ at low temperature. They attributed the low TMR ratio to the imperfect spin filtering in the $\Sigma_{1}$ state. Belashchenko {\it et al.} \cite{2005Belashchenko-PRB} theoretically analyzed Co/Al$_2$O$_3$/Co(111) MTJs. Although they revealed the sensitivity of the current spin polarization to the interfacial atomic configuration, the obtained TMR ratios were quite low ($\sim 60\%$). In the present work, by focusing on simpler (111)-oriented MTJs and carrying out detailed analyses on local electronic structures, we found a quite high TMR ratio and clarified its underlying mechanism.

\section{summary}
We theoretically investigated the TMR effect in unconventional (111)-oriented MTJs, Co/MgO/Co(111) and Ni/MgO/Ni(111). By estimating their transport properties using the first-principles-based approach, we found that the Co-based MTJ has a TMR ratio over 2000\%, which is much higher than that of the Ni-based one of 250\%. The analyses of the LDOSs and the transmittance showed that the obtained high TMR ratio comes from the resonance of the interfacial $d$--$p$ antibonding states in the majority-spin channel. This mechanism is essentially different from that in the Fe(Co)/MgO/Fe(Co)(001) MTJ and suggests a novel way to achieve a giant TMR ratio.

\begin{acknowledgments}
The authors are grateful to Y. Sonobe, S. Mitani, H. Sukegawa, Y. Kozuka, and K. Nawa for useful discussions and helpful comments. This work was partly supported by Grants-in-Aid for Scientific Research (S) (Grant No. 16H06332) from the Ministry of Education, Culture, Sports, Science and Technology, Japan and by NIMS MI$^2$I. The crystal structures were visualized using VESTA \cite{2011Momma-JAC}.
\end{acknowledgments}


\begin{thebibliography}{99}
\bibitem{remark_AP-conductance} In Fig. \ref{kpara-dep}, we only showed $G_{{\rm AP},\uparrow}({\bf k}_\parallel)$ as the conductance in the antiparallel magnetization case, because $G_{{\rm AP},\downarrow}({\bf k}_\parallel)$ has almost the same ${\bf k}_\parallel$ dependence as $G_{{\rm AP},\uparrow}({\bf k}_\parallel)$.
\end{thebibliography}

\end{document}